\documentclass[runningheads]{llncs}
\usepackage[T1]{fontenc}
\usepackage[margin=4cm]{geometry}
\usepackage{algpseudocode}
\usepackage{algorithm}
\usepackage{float}
\usepackage{graphicx}
\usepackage{xcolor}
\begin{document}
\title{SARS: A Resource Selection Algorithm for Autonomous Driving Tasks in Heterogeneous Mobile Edge Computing}

\titlerunning{SARS: A Resource Selection Algorithm for...}
%
%
\author{Reza Zakerian\inst{1}\orcidID{0009-0009-3000-2748} \and
Hadi Gholami\inst{2}\orcidID{0000-0003-4515-9384}}
\authorrunning{R. Zakerian and H. Gholami}
%
\institute{Saint Louis University, Saint Louis, MO, USA 63103 \\
\email{reza.zakerian@slu.edu} \and
Independent Researcher, Amol, Mazandaran, Iran \\
\email{gholamihd@gmail.com}}

\maketitle              
\begin{abstract}
With the rapid advancement of devices requiring intensive computation—such as Internet of Things (IoT) devices, smart sensors, and wearable technology—the computational demands on individual platforms with limited resources have escalated, necessitating the offloading of the generated tasks by the devices to edge. These tasks are often real-time with strict response time requirements. Among these devices, autonomous vehicles present unique challenges due to their critical need for timely and accurate processing to ensure passenger safety. Selecting suitable servers in a heterogeneous mobile edge computing (MEC) architecture is vital to optimizing real-time task processing rates for such applications. To address this, we present an algorithmic solution to improve the allocation of heterogeneous servers to real-time tasks, aiming to maximize the number of processed tasks. By analyzing task and server characteristics in the MEC architecture, we develop the suitability-based adaptive resource selection (SARS) algorithm, which evaluates server suitability based on factors like time constraints and server capabilities. Additionally, we introduce the proactive on-demand resource allocation (PORA) algorithm, which strategically reserves computational resources to ensure availability for critical real-time tasks. We compare the proposed algorithms with several classical and state-of-the-art algorithms. Computational results demonstrate that our approach outperforms existing algorithms, processes more tasks, and effectively prioritizes urgent tasks, particularly in autonomous driving applications.

\end{abstract}
\section{Introduction}
The rapid proliferation of Internet of Things (IoT) devices and the increasing demand for real-time data processing have significantly strained individual platforms with limited computational resources. Devices such as smart sensors, wearable technology, and industrial equipment generate vast amounts of data requiring immediate analysis~\cite{Garikapati2024}. Mobile edge computing (MEC) has emerged as a solution to this challenge, processing data at the network's edge to reduce latency and improve responsiveness~\cite{Chen2022}. However, applications like autonomous vehicles present unique challenges that differ from typical IoT devices due to their safety-critical requirements and the need for instantaneous decision-making based on complex sensor data~\cite{Abdel2024}. Autonomous vehicles must process high volumes of sensor data from radar, LiDAR, and cameras to perceive their environment, make decisions, and compute driving trajectories. Tasks such as object detection, path planning, and sensor fusion are computationally intensive and time-sensitive, with delays potentially leading to catastrophic consequences. Unlike other IoT devices, autonomous vehicles operate in dynamic environments where the distances between vehicles, roadside units (RSUs), and edge computing servers (ECSs) vary, affecting network latency and resource availability~\cite{Dai2019}.

Offloading tasks in autonomous driving is inherently time-critical, and it is imperative to ensure that tasks are completed before their respective deadlines~\cite{Wang2024}. Failure to meet these deadlines renders tasks invalid, compromising the safety of autonomous vehicles and increasing the risk of collisions. Consequently, efficient task offloading is essential to guarantee that all critical tasks are completed within their required time constraints. This process involves two phases: the task selection phase (TSP), which prioritizes tasks, and the resource selection phase (RSP), which assigns tasks to appropriate servers. Both phases are essential for high quality of service, minimizing latency, and ensuring system reliability crucial for real-time decision-making. However, while researchers have developed numerous innovative TSP approaches, they have largely relied on classical RSP algorithms, such as selecting servers based on shortest execution times~\cite{Misra2019,Liu2019} or randomly~\cite{Stankovic1998}. Relying solely on advanced TSP algorithms without improved RSP can result in higher latency, underutilization, and ECS overloading.

When the number of time-constrained tasks on ECSs increases, factors affecting task completion go beyond task selection, network delay, and distances to include how tasks are assigned to handle computationally intensive workloads. To meet deadlines, tasks must be completed on time~\cite{Li2019}, making the RSP algorithm crucial for both ECS service providers and subscribers. Subscribers expect high-quality service, while providers aim to serve as many users as possible. The simplest and most common way to assign a selected task to a server is by assigning it to any server that can process it with the shortest execution and waiting times. However, this is not a feasible solution due to the heterogeneous nature of computing in ECSs, where servers vary in capabilities. Assigning tasks to the server with the shortest execution and waiting times may not yield near-optimal performance, leading to suboptimal resource utilization and increased processing times. For example, assigning a computationally intensive task to a less powerful server could result in delays, even if that server is immediately available. Additionally, an ECS located near a stadium during a baseball match may experience unbalanced resource allocation and diminished quality of service without an efficient resource selection strategy to properly offload tasks onto the ECSs. Similarly, an ECS situated near a highway during rush hours may be overwhelmed by a large volume of traffic, resulting in a lower success rate of requests.

To address the challenge of advancing task offloading from autonomous vehicles to heterogeneous ECSs, aiming to maximize the number of processed tasks, we propose two new algorithms that enable ECSs to process as many constrained tasks as possible. Our primary contributions in this paper are as follows:
\begin{itemize}
    \item We develop a suitability-based adaptive resource selection (SARS) algorithm to maximize the execution of time-constrained tasks by efficiently assigning them to the most appropriate servers. It considers critical factors such as task characteristics, time constraints, and urgency levels to optimize resource utilization and ensure a high task completion rate.
    \item We propose an algorithm called proactive on-demand resource allocation (PORA) to reserve resources in ECSs and proactively allocate them based on real-time demand. PORA ensures that critical tasks with imminent deadlines receive priority, maintaining system responsiveness and efficiency.
    \item We conduct a thorough experimental evaluation demonstrating that SARS and PORA perform more efficiently, exhibiting a significant improvement in the task completion rate metric. Using the task selection approaches but with different RSP algorithms, our approaches yield superior results achieving up to 13\% increase in task completion rate and providing significant benefits for urgent tasks in the tested scenario.
\end{itemize}
The remainder of this paper is organized as follows. Section~\ref{sec:relatedworks} reviews related works. Section~\ref{sec:systemmodel} describes the system design and architecture and formulates the problem. Section~\ref{sec:proposedmethod} details the proposed method. Section~\ref{sec:experimental} presents experimental evaluation and analysis. Section~\ref{sec:conclusion} offers conclusions.

\section{Related Work}
\label{sec:relatedworks}
The emergence of 5G and IoT has heightened interest in using mobile technologies to enhance autonomous driving. These advancements have driven MEC and its task offloading strategies, which are complex due to algorithmic approaches (TSP and RSP) and edge layer architectures (single or multi-server, homogeneous or heterogeneous ECS). This section reviews existing literature on these two aspects.

\subsection{Approaches for TSP and RSP}
\raggedbottom
Several studies have investigated offloading computationally intensive tasks with different RSPs to nearby servers. Misra et al.~\cite{Misra2019} employed a utility function in the local decision-making process for tasks and selected the server with the shortest execution and waiting times for the RSP. Stankovic et al.~\cite{Stankovic1998} employed a random server selection strategy. Xu et al.~\cite{Xu2020} introduced the greedy for energy (GFE) algorithm, assigning tasks to servers that consume the least amount of energy in the system. Similarly, Azizi et al.~\cite{Azizi2022} identified servers that satisfy task deadlines while minimizing energy usage, then randomly assigned tasks to one of the chosen servers. The study~\cite{Gholami2022} after ranking the servers, determined the probability of being selected for the servers with higher rank in order to achieve a better selection. Gholami and Zakerian~\cite{Gholami2020} proposed a concept called emphasized processor, in which in TSP, a server is selected by considering rules, so that in the second phase, it is assigned to the desired server without considering the RSP's algorithm.

Research on TSP in edge computing has been even more extensive. Balasekaran et al.~\cite{Balasekaran2021} introduced the low-complexity earliest hyper period first (EHF) algorithm to prioritize tasks based on urgency and hyper period, improving completion times. Nie et al.~\cite{Nie2023} proposed a multi-agent deep reinforcement learning (MADRL) approach, where tasks are modeled as agents that collaborate through a shared reward function to minimize delays without specific task priorities. Feng et al.~\cite{Feng2017} utilized a decentralized ant colony optimization (ACO) meta-heuristic for scheduling tasks across servers, aiming to optimize job assignments based on computation and resource availability to minimize response times and efficiently distribute workloads.

Other research has focused on prioritizing tasks for the TSP based on different criteria. Liu et al.~\cite{Liu2019} developed an algorithm tailored for autonomous driving tasks, assigning priorities based on task types and dynamically adjusting deadlines to maximize task completion within time constraints. Choudhari et al.~\cite{Choudhari2018} and Dai et al.~\cite{Dai2019} introduced algorithms that categorize tasks into high, medium, and low priorities. Saba et al.~\cite{Saba2021} developed a priority queuing model with preemption to manage tasks, dividing them into critical and non-critical categories and assigning higher priority to time-critical tasks. Yu et al.~\cite{Yu2024} employed dynamic service caching based on task popularity, ensuring frequently requested tasks are readily available. Furthermore, Liu et al.~\cite{Liu2020} and Bini et al.~\cite{Bini2004} focused on deadline-based prioritization and branch-and-bound search methods to maximize system performance without missing deadlines.

\subsection{Edge Computing Architecture}
Edge computing architectures are generally classified into heterogeneous and homogeneous ECSs. In heterogeneous environments, Balasekaran et al.~\cite{Balasekaran2021} introduced a mix of ECSs with varying processing capabilities. Lin et al.~\cite{Lin2024} developed the adaptive priority-based hierarchical task-offloading (APHTO) algorithm for autonomous driving in such settings. Wang et al.~\cite{Wang2018} optimized task scheduling using an ACO algorithm to balance load and reduce energy consumption. Li et al.~\cite{Li2021} created a task offloading strategy that balances task delays and offloading costs, while Zhao et al.~\cite{Zhao2023} implemented a dual-tier edge framework to ensure quality of service (QoS) by categorizing tasks based on latency requirements.

Conversely, homogeneous ECS environments have been addressed by Liu and Chang~\cite{LiuChang2018}, who proposed an interior point method (IPM)-based algorithm to minimize computation delays and energy consumption. Liu and Zhang~\cite{LiuZhang2022} presented a reverse auction-based greedy strategy to reduce offloading costs in homogeneous ECS environments.

The number of ECSs also varies among studies. Rublein et al.~\cite{Rublein2024} designed a two-round bidding task assignment strategy for multi-server systems, incorporating preemption and a greedy heuristic for scalability. Jeremiah et al.~\cite{Jeremiah2024} introduced a digital twin-assisted framework with a deep reinforcement learning algorithm for multi-server vehicular edge computing architectures. In contrast, Guo et al.~\cite{Guo2022} focused on single-server setups, developing a fairness-oriented task offloading strategy using iterative algorithms to optimize task distribution among multiple users.
 
\subsection{Motivation of This Paper}
While studies have explored task scheduling and resource allocation in MEC, they often fall short in addressing the complexities of real-world environments, particularly systems with multiple geographically distant and heterogeneous ECSs. Moreover, efficient resource allocation and the creation of algorithmic strategies to maximize resource utilization remain largely unexplored. In this paper, we develop an RSP algorithm to optimize resource allocation and an algorithm to reserve resources in ECSs, taking into account ECS heterogeneity, varying distances between ECSs and vehicles, and different task types with time constraints. By addressing these factors, our work bridges the gap between theoretical research and practical applications, enhancing the efficiency of heterogeneous ECSs within MEC architectures.

\section{System Model and Problem Formulation}
\label{sec:systemmodel}
\subsection{System Model}
\begin{figure}
\centering
\includegraphics[width=0.7\textwidth, height=5.5cm]{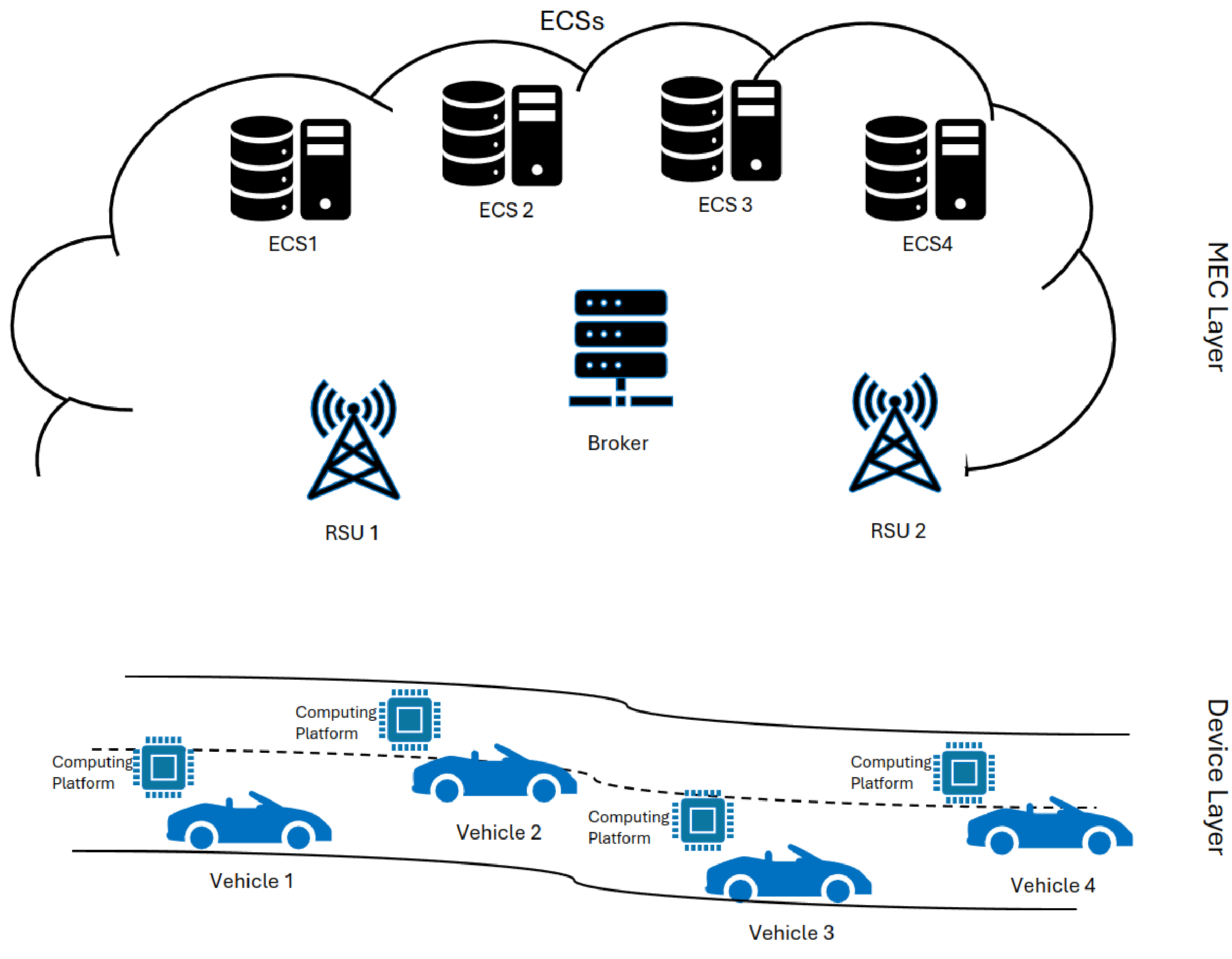}
\caption{An illustration of our system architecture.} \label{edge_env}
\end{figure}

We consider a MEC architecture comprising multiple vehicles, multiple RSUs, a centralized broker, and multiple ECSs. In this architecture, large-scale sensor data of the vehicles need to be processed by ECSs to meet stringent timing requirements via high-speed 5G networks. An overview of the system model is shown in Fig.~\ref{edge_env}.

Our architecture features a two-layer hierarchical network. The \emph{device layer} consists of multiple vehicles at various locations, each equipped with computing platforms that generate tasks from sensor data (e.g., radar, LiDAR, cameras) to perceive their environment. The \emph{MEC layer} includes a set of RSUs that relay tasks to a centralized broker, which handles TSP and RSP by distributing tasks to multiple heterogeneous ECSs based on urgency, priority, and available computational resources. Each ECS contains several heterogeneous processing units (PUs) with different computational capabilities.

Once vehicles initialize and generate tasks, they are first prioritized using the TSP algorithm. The RSP algorithm then assigns each task to an appropriate ECS and specific PU. Tasks can be migrated among PUs within ECSs but become invalid if no PU can execute them. The 5G network facilitates data transmission from vehicles to ECSs within milliseconds, introducing transmission delays for tasks sent from vehicles to the broker and from the broker to ECSs. We account for these transmission delays alongside execution time, which depends on each PU’s processing rate. Additionally, in this real-time system, tasks may experience queue delays in the broker until a PU becomes available for execution.

\subsection{Problem Formulation}
In this system, we consider a set of vehicles denoted as $V = \{v_1, v_2, \dots, v_{vn}\}$, where $vn$ represents the number of vehicles in the device layer. Let us assume $T = \{t_1, t_2, \dots, t_n\}$ is the set of $n$ independent tasks that are generated by the vehicles for processing in the MEC layer. Each task $t_i \in T$ can be generated by a vehicle $v_k \in V$ and is defined by a tuple $t_i = \langle t_i^r, t_i^d, t_i^w, t_i^s \rangle$, where $t_i^r \geq 0$ is the release time, $t_i^d > t_i^r$ is the deadline, $t_i^w$ is the computational workload required by the task in MI (million instructions), and $t_i^s$ is the file size of the task in kilobytes. The interval $[t_i^r, t_i^d]$ defines the time window within which the task $t_i$ must be processed. The set of tasks $T$ should be offloaded to the MEC layer to be processed by the set of ECSs $E = \{ec_1, ec_2, \dots, ec_{en}\}$, where $en$ denotes the number of heterogeneous ECSs. There is a communication link between the two sets $V$ and $E$ to realize this offloading.

Let us consider $L$ as the set of communication links in this two-layer network. We assume two kinds of communication links: the link between vehicles $V$ and RSUs $RS = \{rs_1, rs_2, \dots, rs_{rn}\}$, and the link between the broker $B$ and ECSs $E$, where $rn$ denotes the number of RSUs. As $rs_i \in RS$ and the broker $B$ are close together, this link has a negligible effect on the delay in transferring the tasks, so we skip reporting it. Therefore, the set of communication links can be represented as $L = \{e_{1,rs_1}, e_{2,rs_2}, \dots, e_{vn,rs_{rn}}, e_{B,1}, e_{B,2}, \dots, e_{B,en}\}$. These communication links/connections lead to a delay in a task called the transmission delay. The transmission delay experienced by a task $t_i$ can be represented as:
\begin{equation}
    d_i^{td} = \left( \frac{t_i^s}{bw_{ej,k}} \times dis_{j,k} \right) + \left( \frac{t_i^s}{bw_{e_{B,u}}} \times dis_{B,u} \right)
\end{equation}
where $dis_{j,k}$ is the distance between $v_j \in V$ and $rs_k \in RS$, $bw_{e_{B,u}}$ is the distance between $B$ and $ec_u \in E$, and $bw_{ej,k}$ and $bw_{e_{B,u}}$ are the bandwidths of $e_{j,k}$ and $e_{B,u}$, respectively.

A task \( t_i \in T \) experiences a queue delay \( d_i^{bq} \) after entering the broker \( B \). This delay represents the time the task spends waiting in the broker's queue before being assigned to an available resource.

For each $ec_k \in E$, define $ec_k = \{c_k^1, c_k^2, \dots, c_k^{cn}\}$ as the set of heterogeneous cores (or PUs) that work in parallel. Let the PU $c_k^j \in ec_k$ have a processing rate $rc_k^j > 0$. We assume that a task $t_i \in T$ may be processed on any PU $c_k^j \in ec_k$ without preemption, and interruption is not allowed during the processing. Then, the delay $d_i^{pro}$ for processing $t_i \in T$ on PU $c_k^j \in ec_k$ is calculated as follows:
\begin{equation}
\label{eq:delay}
    d_i^{pro} = \frac{t_i^w}{rc_k^j}
\end{equation}

Therefore, the completion time of $t_i \in T$ can be calculated as follows:
\begin{equation}
    t_i^{ct} = t_i^r + d_i^{td} + d_i^{bq} + d_i^{pro}
\end{equation}

We ensure that a task is considered completed/processed when $t_i^{ct} \leq t_i^d$. Hence, the number of processed tasks $N_{pt}$ is defined as follows:
\begin{equation}
    N_{pt} = \sum_{\forall i \in T} t_i
\end{equation}

The objective function is the maximization of the number of tasks processed and can be denoted as follows:
\begin{equation}
    \max (N_{pt})
\end{equation}

\section{Proposed Method}
\label{sec:proposedmethod}
Due to the characteristics and urgency of autonomous driving tasks, our goal is to solve the resource allocation problem to execute as many tasks as possible before their deadlines, thereby maximizing the number of processed tasks $N_{pt}$. To achieve this, we must efficiently process tasks while considering their stringent time constraints and varying priorities. This involves two key components: a TSP algorithm to decide which task should be selected next from the waiting queue based on factors like deadlines and priorities, and an RSP algorithm to determine to which PU within an ECS the selected task should be assigned. This is where our proposed SARS algorithm comes into play within the RSP. The SARS algorithm evaluates all available PUs across the ECSs to find a suitable PU for processing the task. If no PU can satisfy the time interval $[t_i^r, t_i^d]$, the SARS algorithm engages the PORA algorithm to activate a reserved PU. If successful, the PORA algorithm activates the reserved PU, processes the urgent task, and then returns the PU to standby. If no reserved PU is available, the task is marked as invalid. By integrating a TSP for task selection, the SARS algorithm within the RSP for resource selection, and the PORA algorithm for resource reservation, we ensure that tasks are prioritized, assigned, and executed efficiently.

\subsection{Suitability-based Adaptive Resource Selection Algorithm}
To assign tasks offloaded from vehicles $V$ to PUs within ECSs, we propose the suitability-based adaptive resource selection (SARS) algorithm. SARS aims to assign each task to the most appropriate PU, ensuring that task deadlines are met while balancing the load across all PUs. When tasks are offloaded from vehicles $V$ to the broker $B$, they enter a dynamically managed waiting queue. This queue is rearranged in each time unit using a TSP algorithm, such as the earliest deadline first (EDF), which assigns higher priorities to tasks with earlier deadlines. When a task $t_i$ becomes ready for assignment, the SARS algorithm evaluates all available PUs across the ECSs to select the most suitable one. The pseudo-code of the proposed SARS algorithm is detailed in Algorithm~\ref{alg:SARS}.

As shown in Algorithm~\ref{alg:SARS}, given a task \( t_i \), a set of ECSs \( E \), and, for each ECS \( ec_k \in E \), a set of PUs in \( ec_k \), the SARS algorithm selects the most suitable PU \( { c_k^j }^* \) to process the task \( t_i \). 
To achieve this, the algorithm in for-loop Line 6 iterates over each ECS \( ec_k \) in the set \( E \). For each ECS, it further iterates through each PU \( c_k^j \) within that server in Line 7 to calculate a suitability score for every PU. To compute the suitability score, it incorporates two essential components: the first component is the time margin, calculated using Equation~(\ref{eq:time_margin}) in Line 8, and the second component is the load factor, calculated using Equation~(\ref{eq:load_factor}) in Line 9. Once these two components are calculated, it computes the suitability score based on Equation~(\ref{eq:suitability_score}) in Line 11. This calculated suitability score helps SARS to find the most suitable PU for the given task. In case there is no suitable PU to meet the time constraint of the task, the SARS algorithm invokes the PORA algorithm in Line 22 to make a reserved resource available for the task.
\begin{algorithm}[hbt!]
\caption{The pseudo-code for SARS}\label{alg:SARS}
\begin{algorithmic}[1]
\State \textbf{Input:} {Task $t_i$, set of ECSs $E$, set of PUs $ec_k$}
\State \textbf{Output:} {Suitable PU \( {c_k^j}^* \) or task marked as invalid}
\State \textbf{Initialize Variables}
\State \quad \( {c_k^j}^* \gets {None} \)
\State \quad \( {max\_ss} \gets -\infty \)
\For{each edge server \( ec_k \in E \)}
    \For{each PU \( {c_k^j} \in ec_k \)}
        \State Calculate ${tm}_i^{c_k^j}$ using Equation~(\ref{eq:time_margin})
        \State Calculate ${lf_i^{c_k^j}}$ using Equation~(\ref{eq:load_factor})
        \If{\( \mathrm {tm}_i^{c_k^j} \geq 0 \)}
            \State Calculate ${ss_i^{c_k^j}}$ using Equation~(\ref{eq:suitability_score})
            \If{\( {ss_i^{c_k^j}} > {max\_ss} \)}
                \State \( {max\_ss} \gets {ss_i^{c_k^j}} \)
                \State \( {c_k^j}^* \gets {c_k^j} \)
            \EndIf
        \EndIf
    \EndFor
\EndFor
\If{\( {c_k^j}^* \neq {None} \)}
    \State \textbf{return} {Suitable PU of \( {c_k^j}^* \)}
\Else
    \State Invoke PORA to process \( t_i \) on a reserved PU
    \If{No reserved PU to satisfy interval \([t_i^r, t_i^d]\)}
        \State \textbf{return} {Task \( t_i \) as invalid}
    \EndIf
\EndIf
\end{algorithmic}
\end{algorithm}

We now present the algorithm in more detail. The time margin ${tm}_i^{c_k^j}$ quantifies the spare time available after the estimated completion of task $t_i$ on PU $c_k^j$ before its deadline $t_i^d$. A larger time margin indicates greater flexibility and reduces the risk of missing deadlines. The time margin is defined as:
\begin{equation}
\label{eq:time_margin}
\mathrm{tm}_i^{c_k^j} = t_i^d - t_i^{ct,c_k^j}
\end{equation}
where \( t_i^{ct,c_k^j} \) is the estimated completion time of task \( t_i \) on PU \( c_k^j \) in ECS $ec_k$.

The load factor ${lf_i^{c_k^j}}$ represents the relative load on the PU $c_k^j$ compared to the maximum load among all PUs. It ensures that highly utilized PUs are less likely to be allocated to additional tasks, promoting balanced resource usage. The load factor is calculated as: 
\begin{equation}
\label{eq:load_factor}
\mathrm{lf_i^{c_k^j}} = \frac{ctp^{c_k^j}}{\max {ctp}}
\end{equation}
where $\mathrm{lf_i^{c_k^j}}$is load factor for the PU $c_k^j$, used to assess its relative load. $\mathrm{ctp^{c_k^j}}$ is the current load of the PU $c_k^j$, and $\max ctp$ is maximum current load among all PUs.

After calculating time margin ${tm}_i^{c_k^j}$ and load factor ${lf_i^{c_k^j}}$, we can calculate suitability score of task $t_i$ on PU $c_k^j$ of ECS $ec_k$ as:
\begin{equation}
\label{eq:suitability_score}
{ss_i^{c_k^j}} = t_i^{ct,c_k^j} + \alpha \cdot \mathrm{tm}_i^{c_k^j} + \beta \cdot \mathrm{lf}_i^{c_k^j}
\end{equation}
where ${ss_i^{c_k^j}}$ is the suitability score of $t_i$ on PU ${c_k^j}$ and $\alpha$ and $\beta$ are weighting factors that adjust the importance of the time margin and load factor, respectively. 

Finally, using the previously calculated suitability scores ${ss_i^{c_k^j}}$, the SARS algorithm prioritizes PUs based on their ability to complete tasks within the specified time constraints. For each task $t_i$, the algorithm selects the PU with the highest suitability score, provided it can meet the time interval \([t_i^r, t_i^d]\). If no PU satisfies this time constraint, the task is marked as invalid.

To structure this selection, we formulate the SARS algorithm as follows. For each task \( t_i \in T \), we first identify the subset of PUs that can complete the task by its deadline:
\begin{equation}
ec_k^m = \{ c_k^j \in ec_k \mid \mathrm{tm}_i^{c_k^j} \geq 0 \}
\end{equation}
where \( ec_k \) is the set of all available PUs and the set \( ec_k^m \) includes all PUs \( c_k^j \) that can process task \( t_i \) within its deadline, ensuring that the time margin \( \mathrm{tm}_i^{c_k^j} \) is non-negative.

From the eligible PUs in $ec_k^m$,  we then select the PU with the highest suitability score:
\begin{equation}
{c_k^j}^* = \{ c_k^j \in ec_k \mid {ss_i^{c_k^j}} \geq {ss_i^{c_k^m}} \, \forall c_k^j \in ec_k \}
\end{equation}

This ensures that, among those PUs capable of meeting the task’s timing requirements, the PU with the highest suitability score is chosen to execute the task.

\subsubsection{Proactive On-demand Resource Allocation:}
Given that our primary objective is to maximize the number of processed tasks \( N_{pt} \), we strive to complete as many tasks as possible. However, even with an efficient RSP algorithm, full utilization of all ECSs may not be achieved due to the underutilization of certain PUs. This inefficiency arises because RSP algorithms cannot fully anticipate the characteristics of future real-time tasks; consequently, they may allocate valuable PUs to non-critical tasks, thereby wasting resources that could be better employed for critical workloads. To mitigate this issue, we introduce the proactive on-demand resource allocation (PORA) algorithm into the SARS algorithm.

The core idea of PORA is to prevent the wastage of PUs on non-critical tasks by reserving certain capable PUs in ECSs for computationally intensive tasks with immediate deadlines. By strategically reserving PUs and dynamically allocating them based on real-time demand, PORA enhances resource utilization and ensures that critical tasks receive priority processing. Specifically, when a time-critical task arrives and SARS is unable to assign it to any available PU within the required interval \([t_i^r, t_i^d]\), PORA intervenes by providing a reserved PU to process the task, subsequently returning it to standby for future use. The detailed implementation of the PORA algorithm is presented in Algorithm~\ref{alg:pora}.

\begin{algorithm}[hbt!]
\caption{The pseudo-code for PORA}\label{alg:pora}
\begin{algorithmic}[1]
\State \textbf{Input:} Set of ECSs $E$, set of PUs $ec_k$
\State \textbf{Output:} Reserve PUs
\For{each Edge Computing Server \( ec_k \in E \)}
    \State Calculate average processing rate for \( ec_k \) using Equation~(\ref{eq:average_speed})
    \State Initialize \( diff\_list \gets {None} \)
    \State Initialize \( c_k^{j*} \gets {None} \)
    \For{each PU \( c_k^j \in ec_k \)}
        \State Calculate absolute difference from the average processing rate for PU $c_k^j$ using Equation~(\ref{eq:diff})
        \State Append \((diff, c_k^j)\) to \( diff\_list \)
    \EndFor
    \State Select the top \( k \) PUs with the smallest absolute difference from \( diff\_list \)
    \State Randomly choose one PU from the top \( k \) PUs and assign it to \( c_k^{j*} \)
    \State Mark \( c_k^{j*} \) as standby
\EndFor
\State \textbf{return} {Standby PUs}
\end{algorithmic}
\end{algorithm}

The algorithm starts by taking as input a set of ECSs \( E \); for each ECS \( ec_k \in E \), there is a set of PUs. The expected output is a set of PUs to be reserved. The outer for-loop in Line 3 serves as the main body of this algorithm, where it iterates through the set of ECSs \( E \) and calculates the average processing rate of each ECS \( ec_k \) ($\bar{a_k}$) in Line 4 using the following equation:
\begin{equation}
\label{eq:average_speed}
\bar{a_k} = \frac{1}{M} \sum_{m=1}^{M} rc_k^m
\end{equation}
where \( M \) is the number of PUs in ECS \( ec_k \), and \( rc_k^m \) denotes the processing rate of the PU \( c_k^m \) in ECS \( ec_k \).

Then, in Line 5, the algorithm initializes a list \( diff\_list \) to store the absolute differences between the processing rates of each PU \( c_k^j \) and the average processing rate \( \bar{a}_k \). It also initializes a variable \( c_k^{j*} \) in Line 6 to hold the selected PU for each ECS \( ec_k \). The inner for-loop starting at Line 7 iterates over each PU \( c_k^j \) in \( ec_k \). In Line 8, it calculates the absolute difference between the processing rate of PU \( c_k^j \) and the average processing rate \( \bar{a}_k \) using the following equation:
\begin{equation}
\label{eq:diff}
{diff}_{k,j} = \left| rc_k^m - \bar{a}_k \right|
\end{equation}
This calculated difference is then appended to \( diff\_list \) for that ECS.

After calculating the absolute differences for all PUs \( c_k^j \) in each ECS \( ec_k \), the PUs are ranked based on their \( \mathrm{diff}_{k,j} \) values. In Line 11, the algorithm selects the top \( k \) PUs with the smallest absolute differences. Then, in Line 12, one PU is randomly chosen from these top \( k \) candidates. The selected PU \( c_k^{j*} \) is placed on standby in Line 13, reserving it for critical tasks as needed.

\section{Experimental Results}
\label{sec:experimental}
\subsection{Simulation Setup}

\subsubsection{Environment:}To demonstrate the applicability of our proposed algorithms, we consider a simulation scenario involving several autonomous vehicles operating within a \(0.7\, {km} \times 0.7\, {km}\) area. The distances between vehicles, RSUs, the broker, and ECSs in the 5G network range from 50 meters to 250 meters. The autonomous vehicles continuously generate tasks with varying levels of urgency and seek to offload them to the ECSs. Typically, the closest and fastest ECS tends to become overloaded with tasks. To address this issue, our proposed algorithms apply weighting mechanisms and distribution policies to assign tasks more efficiently and fairly, ensuring that a greater number of tasks are executed. In our simulation, we initialize four vehicles, two RSUs, one broker, and four heterogeneous ECSs. Each ECS contains 8 to 12 heterogeneous PUs with processing rates ranging from 0.5 to 1.2 MHz. Moreover, we tuned the parameter \(\alpha\) in the range of 0.5 to 1.5 and set \(\beta\) to 0.5. Through the trial-and-error method, we found that these parameters effectively weight the calculation of the suitability score in the SARS algorithm.

\subsubsection{Real-time Tasks:}
The experiment consists of four groups of tasks, each comprising 200 tasks, resulting in a total of 800 tasks released from four vehicles to the MEC layer. The tasks within each group differ from those in other groups based on characteristics such as deadlines and release times. The range for the required computational workload of the tasks \(t_i^w\), and the range for their release time \(t_i^r\) and deadline \(t_i^d\) are given in Table~\ref{task_attributes}. In the first group, both release times and deadlines are tight. The second group features loose release times but tight deadlines. The third group has tight release times and loose deadlines, whereas in the fourth group, both release times and deadlines are loose. The required computational workloads for the tasks were randomly selected within a range of 1 to 10 MI. Similarly, release times and file sizes were randomly generated within the ranges specified in Table~\ref{task_attributes}. The deadlines are calculated as the sum of the release time, the required computational workload, and a random number within a specific range; for instance, \(t_i^d = t_i^r + t_i^w + {rand(range)}\).
\begin{table}[h]
\centering
\caption{Real-time tasks attribute settings in the simulation.}
\label{task_attributes}
\renewcommand{\arraystretch}{1.3}
\setlength{\tabcolsep}{12pt}
\begin{tabular}{|c|c|c|c|c|}
\hline
Task Group & \(t_i^r\) & \(t_i^d\) & \(t_i^w\) & \(t_i^s\) \\
\hline
1 & [0, 40] & [1, 10] & [1, 10] & [1, 5] \\
2 & [0, 70] & [1, 10] & [1, 10] & [1, 5] \\
3 & [0, 40] & [1, 20] & [1, 10] & [1, 5] \\
4 & [0, 70] & [1, 20] & [1, 10] & [1, 5] \\
\hline
\end{tabular}
\end{table}
\subsubsection{Baseline Algorithms:}
Several baseline algorithms for both TSP and RSP were used to evaluate the proposed approach. For RSP, we employed selecting the PU with the shortest execution time~\cite{Misra2019}, randomly selecting a PU~\cite{Stankovic1998}, and choosing the PU that processes the task as late as possible while still meeting the deadline~\cite{Azizi2022}. For TSP, we used classical algorithms: first come first serve (FCFS), earliest deadline first (EDF), earliest due date first (EDD), earliest feasible deadline first (EFDF), critical ratio (CR), and cost over time (COVERT). We also compared our algorithms to two customized priority-based scheduling methods: efficient resource allocation (ERA)~\cite{Choudhari2018} and a priority queuing model~\cite{Saba2021}, both adapted for our problem.

\subsubsection{Metric:}
We compared the algorithms using the task completion rate (TCR) metric, calculated as \(TCR = \frac{N_{pt}}{|T|} \times 100\), where $|T|$ is the total number of tasks in the set of \(T\).

\subsection{Comparison to Baseline Algorithms}
\begin{figure}[h]
\centering
\includegraphics[width=0.55\textwidth, height=4.5cm]{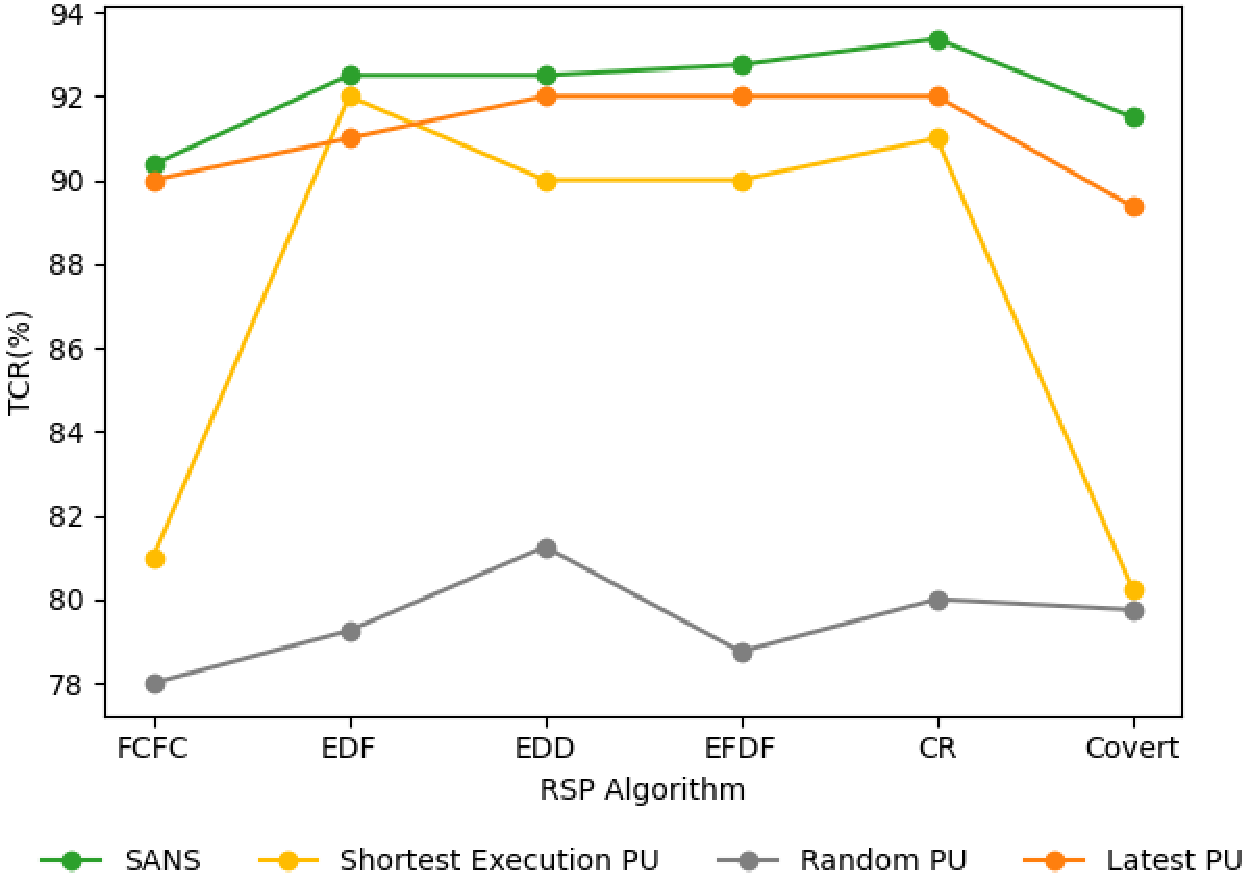}
\caption{The TCR of SARS without PORA compared to other RSP algorithms.}
\label{SARS}
\end{figure}

To evaluate the performance of the proposed algorithm, we executed all TSP algorithms for task selection in combination with all RSP algorithms for resource selection. Subsequently, we executed SARS with all TSP algorithms and compared the results. The comparison results are presented in Fig.~\ref{SARS}, which illustrates the TCR for all TSP algorithms using SARS (in which PORA is disabled) compared to other RSP algorithms. The figure demonstrates that SARS consistently improved the performance of every TSP algorithm and outperformed all other RSP algorithms.

The difference in TCR values between SARS and other RSP algorithms is noteworthy. The best among the other algorithms is the one that selects the PU capable of finishing the task latest while still meeting the deadline. Despite this, it still underperforms compared to SARS across all TSP algorithms, showing a 1.37\% lower TCR in the CR and a 1.5\% lower TCR in EDF. The SARS algorithm demonstrates more stable results across the TSP algorithms due to its suitability-based weighting mechanism. Its performance remains consistently high in all TSP algorithms, while other RSP algorithms, such as the shortest execution PU, exhibit greater fluctuation and less stability. This experiment underscores that SARS not only works better than other resource selection algorithms but is also more reliable, thus achieving superior performance. Notably, this result is achieved solely with SARS, without incorporating PORA.

\begin{figure}[H]
\centering
\includegraphics[width=0.55\textwidth, height=4.5cm]{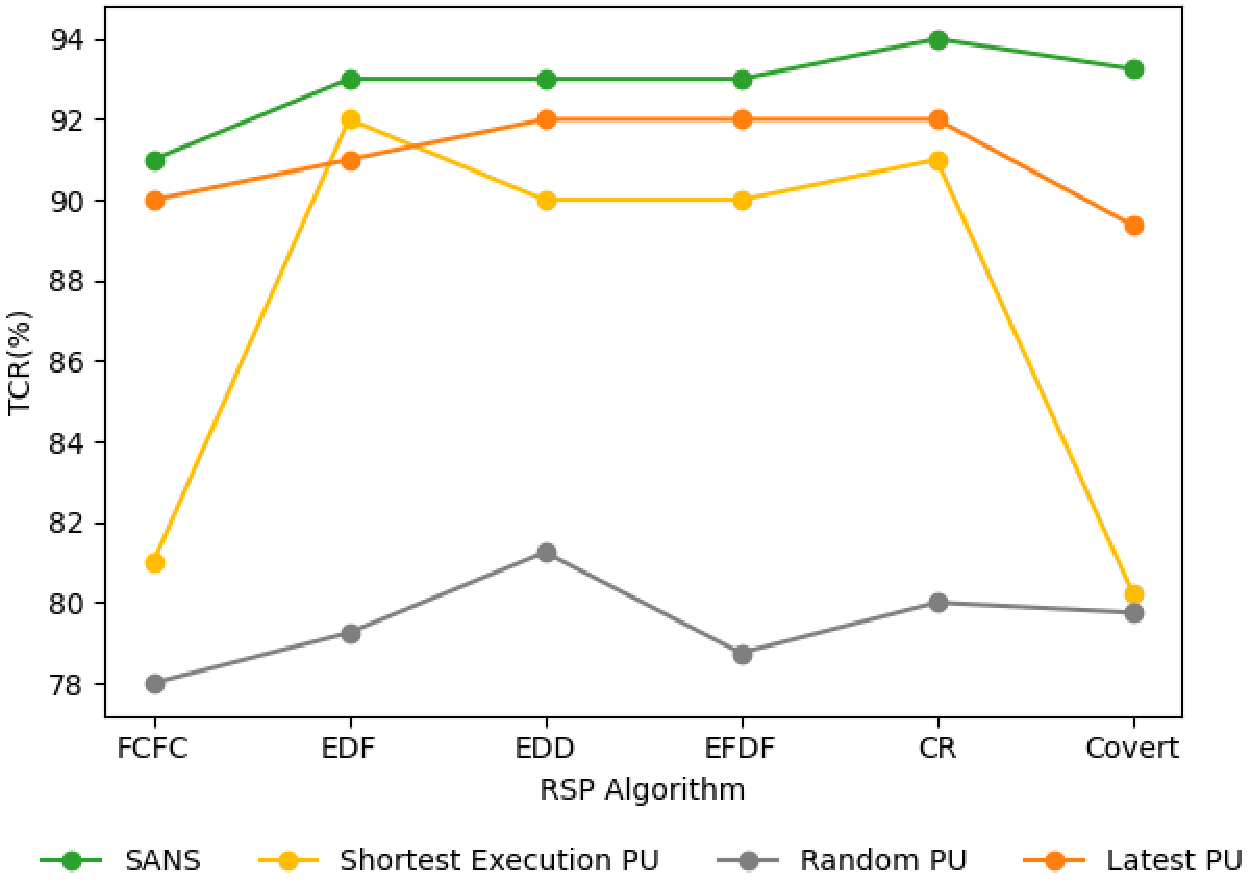}
\caption{The TCR comparison of SARS to other RSP algorithms.}
\label{SARSPORA}
\end{figure}

Fig.~\ref{SARSPORA} compares SARS (in which PORA is enabled) to other RSP algorithms. The enablement of PORA enhances SARS's ability to effectively distribute tasks to PUs while reserving resources for time-critical tasks. Consequently, this leads to a higher TCR, achieving significant improvements in the COVERT algorithm. Specifically, SARS achieves a TCR that is 3.88\% higher than the latest PU algorithm and 13\% higher than the shortest execution PU algorithm. This demonstrates that the integration of SARS and PORA not only benefits the assignment of urgent tasks but also ensures the computation of a greater number of tasks overall.

\begin{figure}[h]
\centering
\includegraphics[width=0.55\textwidth, height=4.5cm]{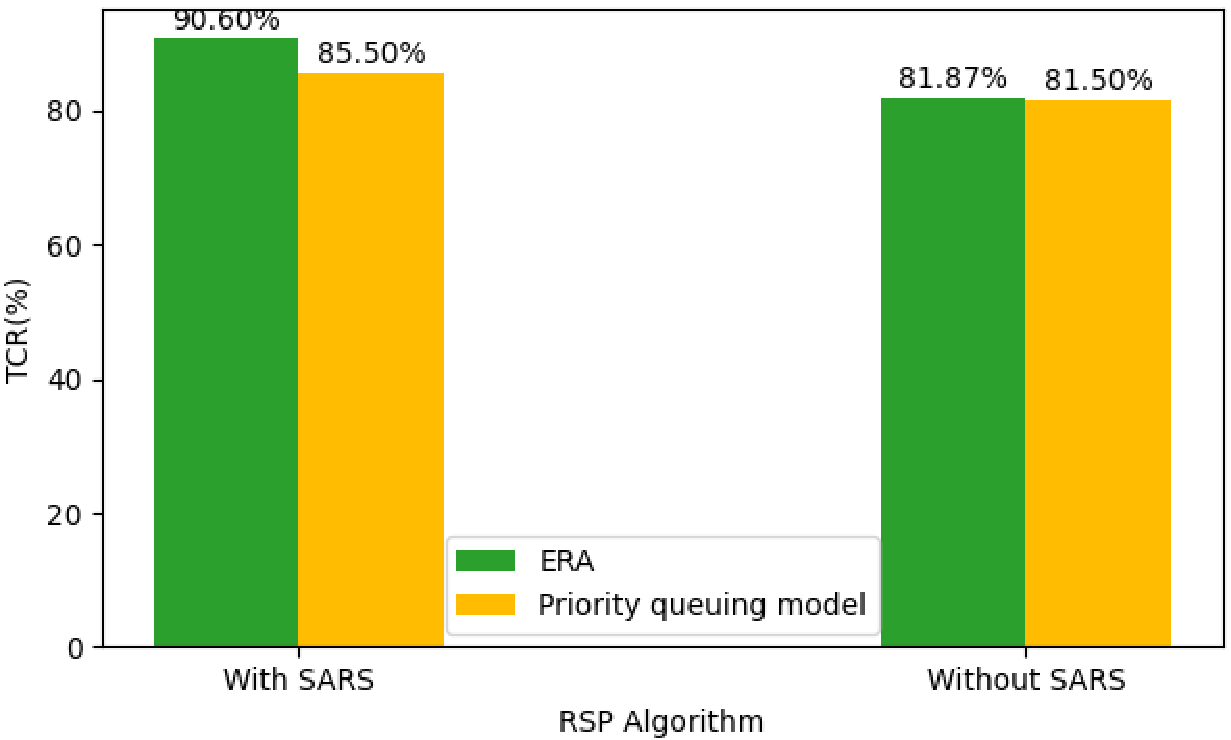}
\caption{Comparison of TCR for the two priority-based algorithms with and without SARS.}
\label{ERA}
\end{figure}

Figure~\ref{ERA} illustrates the impact of SARS as the RSP algorithm on two priority-based scheduling algorithms: ERA and the priority queuing model. We conducted experiments with these priority-based TSP algorithms in two configurations: first, using their original algorithms for both TSP and RSP; second, using their TSP algorithms with SARS as the RSP algorithm. The results indicate that the TCR of ERA and priority queuing model algorithms were improved by 8.73\% and 4.02\%, respectively. These findings highlight the effectiveness and robustness of SARS across diverse task scenarios.

\section{Conclusion}
\label{sec:conclusion}
Given the unbalanced assignment of tasks across heterogeneous ECSs and their PUs, resource allocation for task processing becomes a critical issue. Autonomous vehicles need to offload their real-time tasks to ECSs due to limitations in battery life and computational resources. Moreover, there is a lack of sufficient research on resource allocation in this environment. Therefore, we propose the SARS algorithm to enable heterogeneous ECSs to execute as many tasks as possible based on task characteristics such as time constraints, urgency differences, and the capability of PUs. The SARS algorithm incorporates the idea of resource reservation through the PORA algorithm to reserve resources and prevent computational waste when they are not needed, thereby providing reserved resources for critical tasks. Experimental results demonstrate that SARS significantly improves the TCR across all classical and commonly used TSP algorithms, achieving up to a 13\% increase in performance. Similarly, SARS enhances the performance of priority-based baseline scheduling algorithms by up to 8.73\%.
Although we set the parameters \( \alpha \) and \( \beta \) statically in the suitability score, future work could involve integrating machine learning techniques to dynamically adjust these parameters, enhancing adaptability in various scenarios. Additionally, evaluating the algorithms in larger, more complex environments with multiple brokers, more ECSs, and higher vehicle density would further demonstrate their scalability and robustness.

%
%
%

\begin{thebibliography}{1}
\bibitem{Garikapati2024}
Garikapati, D., Shetiya, S.S.: Autonomous Vehicles: Evolution of Artificial Intelligence and the Current Industry Landscape. Big Data and Cognitive Computing \textbf{8}(4), Article 42 (2024)

\bibitem{Chen2022}
Chen, C., Guo, R., Zhang, W., Yang, J., Yeo, C.K.: Optimal sequential relay-remote selection and computation offloading in mobile edge computing. Journal of Supercomputing \textbf{78}(1), 1093--1116 (2022)

\bibitem{Abdel2024}
Abdel-Aty, M., Ding, S.: A matched case-control analysis of autonomous vs human-driven vehicle accidents. Nature Communications \textbf{15}(1), Article 4931 (2024)

\bibitem{Dai2019}
Dai, H., Zeng, X., Yu, Z., Wang, T.: A scheduling algorithm for autonomous driving tasks on mobile edge computing servers. Journal of Systems Architecture \textbf{94}, 14--23 (2019)

\bibitem{Wang2024}
Wang, Z., Zhao, W., Hu, P., Zhang, X., Liu, L., Fang, C., Sun, Y.: UAV-Assisted Mobile Edge Computing: Dynamic Trajectory Design and Resource Allocation. Sensors \textbf{24}(12), Article 3948 (2024)

\bibitem{Misra2019}
Misra, S., Saha, N.: Detour: Dynamic task offloading in software-defined fog for IoT applications. IEEE Journal on Selected Areas in Communications \textbf{37}(5), 1159--1166 (2019)

\bibitem{Liu2019}
Liu, Q., Chen, Z., Wu, J., Deng, Y., Liu, K., Wang, L.: An efficient task scheduling strategy utilizing mobile edge computing in autonomous driving environment. Electronics \textbf{8}(11), Article 1221 (2019)

\bibitem{Stankovic1998}
Stankovic, J.A., Spuri, M., Ramamritham, K., Buttazzo, G.: Deadline scheduling for real-time systems: EDF and related algorithms. Springer Science \& Business Media, Vol. 460 (1998)

\bibitem{Li2019}
Li, C., Bai, J., Tang, J.: Joint optimization of data placement and scheduling for improving user experience in edge computing. Journal of Parallel and Distributed Computing \textbf{125}, 93--105 (2019)

\bibitem{Xu2020}
Xu, J., Sun, X., Zhang, R., Liang, H., Duan, Q.: Fog-cloud task scheduling of energy consumption optimization with deadline consideration. International Journal of Internet Manufacturing and Services \textbf{7}(4), 375--392 (2020)

\bibitem{Azizi2022}
Azizi, S., Shojafar, M., Abawajy, J., Buyya, R.: Deadline-aware and energy-efficient IoT task scheduling in fog computing systems: A semi-greedy approach. Journal of Network and Computer Applications \textbf{204}, Article 103333 (2022)

\bibitem{Gholami2022}
Gholami, H., Rezvan, M.T.: A cooperative multi-agent offline learning algorithm to scheduling IoT workflows in the cloud computing environment. Concurrency and Computation: Practice and Experience \textbf{34}(22), Article e7148 (2022)

\bibitem{Gholami2020}
Gholami, H., Zakerian, R.: A list-based heuristic algorithm for static task scheduling in heterogeneous distributed computing systems. In: 2020 6th International Conference on Web Research (ICWR), pp. 21--26. IEEE (2020)

\bibitem{Balasekaran2021}
Balasekaran, G., Jayakumar, S., Pérez de Prado, R.: An intelligent task scheduling mechanism for autonomous vehicles via deep learning. Energies \textbf{14}(6), Article 1788 (2021)

\bibitem{Nie2023}
Nie, X., Yan, Y., Zhou, T., Chen, X., Zhang, D.: A Delay-Optimal Task Scheduling Strategy for Vehicle Edge Computing Based on the Multi-Agent Deep Reinforcement Learning Approach. Electronics \textbf{12}(7), Article 1655 (2023)

\bibitem{Feng2017}
Feng, J., Liu, Z., Wu, C., Ji, Y.: AVE: Autonomous vehicular edge computing framework with ACO-based scheduling. IEEE Transactions on Vehicular Technology \textbf{66}(12), 10660--10675 (2017)

\bibitem{Choudhari2018}
Choudhari, T., Moh, M., Moh, T.-S.: Prioritized task scheduling in fog computing. In: Proceedings of the ACMSE 2018 Conference, pp. 1--8. ACM (2018)

\bibitem{Saba2021}
Saba, U.K., Islam, S.u., Ijaz, H., Rodrigues, J.J.P.C., Gani, A., Munir, K.: Planning Fog networks for time-critical IoT requests. Computer Communications \textbf{172}, 75--83 (2021)

\bibitem{Yu2024}
Yu, T.: Optimizing Computational Efficiency in Autonomous Vehicles: Integrative Edge and Cloud Computing Strategies in Vehicular Networks. In: Proceedings of the 2024 11th International Conference on Wireless Communication and Sensor Networks, pp. 5--12 (2024)

\bibitem{Liu2020}
Liu, Y., Wang, S., Zhao, Q., Du, S., Zhou, A., Ma, X., Yang, F.: Dependency-aware task scheduling in vehicular edge computing. IEEE Internet of Things Journal \textbf{7}(6), 4961--4971 (2020)

\bibitem{Bini2004}
Bini, E., Buttazzo, G.C.: Schedulability analysis of periodic fixed priority systems. IEEE Transactions on Computers \textbf{53}, 1462--1473 (2004)

\bibitem{Lin2024}
Lin, J., Rao, H., Liang, S., Zhao, Y., Ren, Q., Jia, G.: Aphto: a task offloading strategy for autonomous driving under mobile edge. The Journal of Supercomputing, 1--33 (2024)

\bibitem{Wang2018}
Wang, T., Wei, X., Tang, C., Fan, J.: Efficient multi-tasks scheduling algorithm in mobile cloud computing with time constraints. Peer-to-Peer Networking and Applications \textbf{11}, 793--807 (2018)

\bibitem{Li2021}
Li, W., Jin, S.: Performance evaluation and optimization of a task offloading strategy on the mobile edge computing with edge heterogeneity. The Journal of Supercomputing \textbf{77}(11), 12486--12507 (2021)

\bibitem{Zhao2023}
Zhao, H., Geng, J., Jin, S.: Performance research on a task offloading strategy in a two-tier edge structure-based MEC system. The Journal of Supercomputing \textbf{79}(9), 10139--10177 (2023)

\bibitem{LiuChang2018}
Liu, L., Chang, Z., Guo, X., Mao, S., Ristaniemi, T.: Multiobjective optimization for computation offloading in fog computing. IEEE Internet of Things Journal \textbf{5}(1), 283--294 (2018)

\bibitem{LiuZhang2022}
Liu, J., Zhang, X.: Truthful resource trading for dependent task offloading in heterogeneous edge computing. Future Generation Computer Systems \textbf{133}, 228--239 (2022)

\bibitem{Rublein2024}
Rublein, C., Mehmeti, F., Mahon, M., La Porta, T.F.: Improved methods of task assignment and resource allocation with preemption in edge computing systems. arXiv preprint \texttt{arXiv:2403.15665} (2024)

\bibitem{Jeremiah2024}
Jeremiah, S.R., Yang, L.T., Park, J.H.: Digital twin-assisted resource allocation framework based on edge collaboration for vehicular edge computing. Future Generation Computer Systems \textbf{150}, 243--254 (2024)

\bibitem{Guo2022}
Guo, K., Zhang, R.: Fairness-oriented computation offloading for cloud-assisted edge computing. Future Generation Computer Systems \textbf{128}, 132--141 (2022)


\end{thebibliography}
%

\end{document}